# Space Subdivision to Speed-up Convex Hull Construction in E³

Vaclav Skala, Zuzana Majdisova, Michal Smolik

Faculty of Applied Sciences, University of West Bohemia,
Univerzitni 8, CZ 30614 Plzen, Czech Republic

**Abstract.** Convex hulls are fundamental geometric tools used in a number of algorithms. This paper presents a fast, simple to implement and robust Smart Convex Hull (S-CH) algorithm for computing the convex hull of a set of points in $E^3$. This algorithm is based on "spherical" space subdivision. The main idea of the S-CH algorithm is to eliminate as many input points as possible before the convex hull construction. The experimental results show that only a very small number of points is used for the final convex hull calculation. Experiments made also proved that the proposed S-CH algorithm achieves a better time complexity in comparison with other algorithms in $E^3$.

## 1 Introduction

A convex hull is a fundamental construction not only in computational geometry and mathematics. It has numerous applications in various fields such as collision detection, mesh generation, shape analysis, cluster analysis, metallurgy, crystallography, cartography, image processing, sphere packing and point location. There are many other problems which can be reduced to the convex hull, e.g. halfspace intersection, Delaunay triangulation, Voronoi diagram, etc. Fast convex hull algorithms are useful for interactive applications, such as collision detection in computer games and path planning for robotics in dynamic environments.

A subset $S \subseteq \mathbb{R}^3$ is convex if and only if for any two points $\boldsymbol{p}, \boldsymbol{q} \in S$ the line segment with endpoints $\boldsymbol{p}$ and $\boldsymbol{q}$ is contained in $S$. The convex hull $\mathcal{CH}(S)$ of a set $S$ is the smallest convex set containing $S$. The convex hull of a set of points $P$ is a convex polyhedron with vertices in $P$.

Many algorithms for calculation of the convex hull in $3D$ have been developed over the last several decades. Chand and Kapur [1] developed the Gift Wrapping algorithm, and Preparata and Hong [2] developed a recursive algorithm, which is based on Divide & Conquer. Clarkson and Shor [3] introduced an incremental insertion algorithm, where the points are processed one by one with respect to the currently constructed convex hull. Barber et al. [4] developed an efficient convex hull algorithm, which is called QuickHull. The time complexity of some of the convex hull algorithms is presented in Table 1.

Several parallel algorithms for convex hull construction were proposed. Chow [12] presented a parallel convex hull algorithm that runs at $O(\log^3 n)$ time complexity.

Amato and Preparata [13] designed an $O(\log^2 n)$ time algorithm using $n$ processors, where $n$ is the number of input points. Reif and Sen [14] proposed a randomized algorithm for three dimensional convex hulls that runs at $O(\log n)$ time using a divide and conquer approach on $O(n)$ processors. Amato et al. [15] gave a deterministic $O(\log^3 n)$ time algorithm for a convex hull in $R^d$ using $O(n \log n + n^{\lfloor d/2 \rfloor})$ work. Gupta and Sen [11] proposed a fast parallel convex hull algorithm that is output-size sensitive.

There are several convex hull algorithms modified for GPU applications. Gao et al. [17] developed a two-phase convex hull algorithm in three dimensions that runs on the GPU. Stein et al. [10] proposed a parallel algorithm based on QuickHull approach.

Other algorithms are based on a probabilistic approach [18]. Precision of convex hull algorithm with regard to physical floating point representation is solved as well, e.g. in [16].

**Table 1.** Comparison of 3D convex hull algorithms and their time complexity. The number of input points is $n$ and $h$ is the number of points on the output convex hull. Note that $h < n$, so $nh < n^2$, usually $h \ll n$.

| Algorithm | Time complexity | Reference |
|---|---|---|
| Gift Wrapping | $O(nh)$ | [1],[7] |
| QuickHull | $O(n \log n)$ | [4] |
| Divide & Conquer | $O(n \log n)$ | [2] |
| Randomized Incremental | $O(n \log n)$ | [3],[5] |
| Chan's algorithm | $O(n \log h)$ | [6] |

## 2   Proposed Algorithm

In this section, we introduce a new Smart Convex Hull (S-CH) algorithm based on space subdivision for construction of the convex hull in E³. The main idea of this algorithm is to eliminate as many input points as possible using an algorithm with $O(N)$ complexity based on space subdivision, and a "standard" convex hull algorithm with $O(n \log n)$ is used for the remaining points, where $n \ll N$. We use "spherical" space subdivision based on $3D$ sectors for efficient elimination of points not contributing to the final convex hull.

This section is organized as follows. Section 2.1 presents the first step of the S-CH algorithm, which is an inner convex polyhedron construction followed by the location of points inside the initial convex polyhedron. In Section 2.2, we describe how to perform the division of points into non-overlapping 3D pyramidal shape sectors. Section 0 presents reduction of the suspicious points. The calculation of a convex hull from the selected points with a standard convex hull algorithm is made in Section 2.4.

### 2.1   Location of Points inside Polyhedron

At the beginning of the proposed S-CH algorithm, we need to find the extremal points in all axes, i.e. points with maximum and minimum $x$, $y$ or $z$ coordinates. The time complexity of this step is $O(N)$. For our purpose, we do not need the exact extremal points, because extremal points close enough are sufficient. This means that we do not have to search extremes through all the input points, but we can search only random sample points. According to experiments made, approx. 10% of all points is sufficient. This simplification does not cause any problems for future calculations and we save computational time as well and the complexity of this step is $O(N)$ only. Therefore, we generally get six distinct extremal points or less.

Now we can create a convex polyhedron using these points, see Fig. 1. Note that the extremal points are determined using the above presented estimation. We assume that the volume of the final object is nonzero, so the convex polyhedron will not be degenerated. One very important property of this polyhedron is that any point lying inside cannot be a point on the convex hull. Thus, we can perform a fast and simple initial test for a point inside/outside the polyhedron and discard many points.

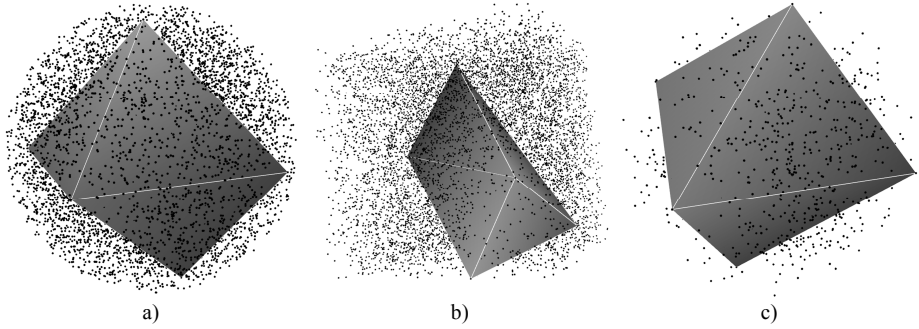

a)  b)  c)

**Fig. 1.** Location of an initial inner testing polyhedron inside the convex hull for $10^4$ points: a) uniform points in sphere, b) uniform points in cube, c) Gauss points.

The location test of a point inside a polyhedron can be performed as follows. Each face of the polyhedron is an oriented plane with a normal vector oriented outside of the polyhedron. Then we can calculate:

$$F_i(\boldsymbol{x}) = a_i x + b_i y + c_i z + d_i = \boldsymbol{n}_i^T \boldsymbol{x} + d_i = 0, \qquad (1)$$

where $\boldsymbol{x}$ is a point and $F_i(\boldsymbol{x}) = 0$ is the implicit equation of a plane with index $i$ having the normal vector $\boldsymbol{n}_i = (a_i, b_i, c_i)$. If $F_i(\boldsymbol{x}) < 0$ for at least one $i \in \{0,1,\dots,7\}$, then point $\boldsymbol{x}$ lies outside of the polyhedron and has to be further processed. Otherwise, point $\boldsymbol{x}$ lies inside of the polyhedron and can be eliminated.

### 2.2 Division of Points into 3D Sectors

In the second step of the S-CH algorithm, only the points, which lie outside of the initial polyhedron, will be further processed. Firstly, we perform the division of $3D$ space into several non-overlapping "pyramidal shape" sectors, i.e. we are using an "approximated spherical" subdivision. A center point and both angles (azimuth $\varphi$ and zenith $\theta$) are used in this subdivision. The center point $C$ is defined as the average of all vertices of the initial polyhedron.

Division of space can be performed as a uniform spherical subdivision in both angles, where azimuth $\varphi \in [0, 2\pi)$ and zenith $\theta \in [0, \pi]$. However, using this, we would have to calculate the exact angles and, moreover, an explosion of small and singular triangles would occur at the both poles. Therefore, we use a simplified calculation of approximated angle. As a result of this simplification, the sectors are not uniformly distributed in the spherical coordinate space, but are uniformly distributed on the faces of a cube, see Fig. 2. Now, when calculating the azimuth and zenith, we have to locate the exact third of the octant, where the point is located and then calculate the intersection with the given face. Calculation of the intersection is easy, because all faces are axes aligned, i.e. $x = \pm 1$ or $y = \pm 1$ or $z = \pm 1$. Finally, we have to determine a table of neighbors for each sector. Note that the neighboring sector can lie on another face of the cube. This means that adjacency of sectors can be determined across the edge of a cube or the vertex of a cube.

Now we are able to calculate the exact index of a sector to which the given point belongs.

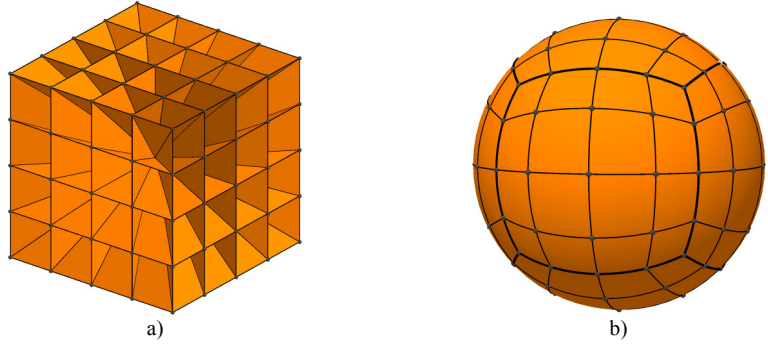

**Fig. 2.** Division of space into 96 (16 × 6 faces) non-overlapping sectors uniformly distributed on a cube: a) sectors displayed on a cube, b) sectors displayed on a sphere.

For each sector with index $i$, one maximal point $R_i^{max}$ is determined. This point equals a point where is a maximum distance between the center point $C$ and all points in a sector. The initial points $R_i^{max}$ are lying on the faces of the initial polyhedron. These points can be calculated as an intersection point of the axis of a sector and the face of the initial polyhedron.

For each new point we have to check whether the distance from this point to the center point $C$ is greater than the distance from $R_i^{max}$ to the center point $C$. If this is true, then we have to replace point $R_i^{max}$ with a processed point, add this point into the sector with index $i$ and recalculate the test planes, see Fig. 3. Otherwise we continue with the next step.

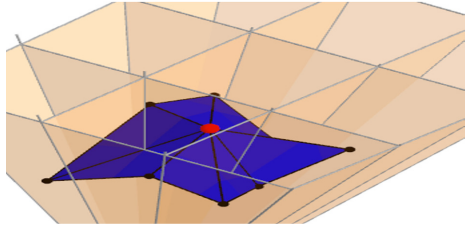

**Fig. 3.** Visualization of testing planes.

In the next step, we check whether the processed point lies over or under the test planes.

Firstly, we determine the projection of the actual point to the face of the unit cube. Then we can compare coordinates of this projection with the projection of maximal point $R_i^{max}$ and based on the result, we choose one of four options, see Fig. 4. Now we have to use the five planes which are defined by maximal points $R_i^{max}$ of the actual sector and neighboring sectors (hatched green) and perform a test for a point over/under the plane. If the point is under all five planes, we can discard it, because such a point cannot be part of the convex hull. Otherwise we add this point into the sector with index $i$.

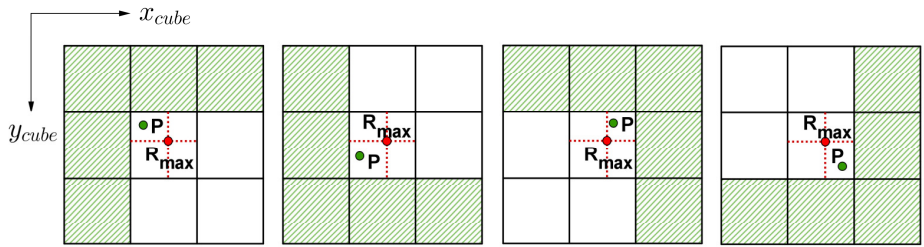

**Fig. 4.** Schema of a testing point with respect to the test planes.

We can gain some extra speed-up if the input dataset is pre-sorted according to the distance from the center point $C$. In such case we start by processing the farthest points from the input dataset. It leads to fast determination of maximal points $R_i^{max}$ and more points from the input dataset can be eliminated. Moreover, the next step, which is described in Section 2.3, does not need be performed.

The pre-sorted input dataset can speed-up the reduction steps. However, the sorting algorithms have the time complexity $\mathcal{O}(N \log N)$, which is higher than the time complexity of reduction steps $\mathcal{O}(N)$. Therefore, it is not beneficial to sort the input dataset.

## 2.3 Reduction of Suspicious Points

We have already divided all suspicious points into sectors. We gave points $R_i^{max}$ some initial values before starting to divide points into non-overlapping sectors and we used these points $R_i^{max}$ to check whether to add or eliminate a point. Values of points $R_i^{max}$ have changed during the division process; hence we have to recheck all suspicious points using the final values of points $R_i^{max}$. We minimize the number of suspicious points, which are input for the final convex hull construction, using this step. Final sets of suspicious points for input datasets with different distributions of points are shown in Fig. 5.

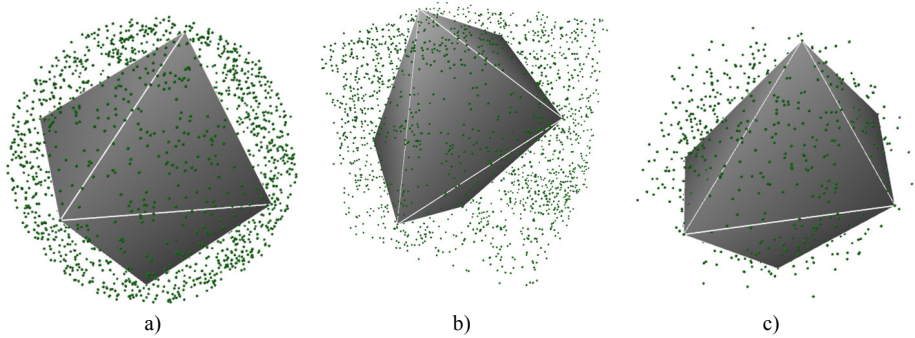

a)          b)          c)

**Fig. 5.** Suspicious points that are the input for convex hull creation ($10^4$ input points): a) uniform points in sphere, b) uniform points in cube, c) Gauss points.

It should be noted that the reduction test eliminates the vast majority of given points. In case that the majority of points are close to the surface of the corresponding convex hull then the performance of reduction steps will decreases as only few points will be reduced.

## 2.4 Convex Hull Construction

After performing the previous steps, we use any known algorithm for calculation of the convex hull. The set of input points for this algorithm equals suspicious points. The number of suspicious points is extremely low in comparison of the number of the original points; thus the time needed for determining the convex hull is insignificant compared to the time needed for reduction of the original input points. Therefore, this step is more or less independent of the choice of a convex hull algorithm. In our approach we used the library MIConvexHull[1], which is based on the QuickHull algorithm.

QuickHull uses a divide and conquer approach. This algorithm performs the following steps:

---

[1] This library is available at https://miconvexhull.codeplex.com/.

1. Find three points (for example, points with minimum and maximum $x$ coordinates and a point with a minimum $y$ coordinate) which are bound to be part of the convex hull.
2. Divide the set into two subsets of points by a plane formed by the three points. This step will be processed recursively.
3. On one side of the plane, determine the point with the maximum distance from the plane. The three points found before along with this one form a pyramid.
4. In the next step, the points lying inside of the pyramid can be ignored.
5. Repeat the previous two steps on the three planes formed by the pyramid.
6. Repeat this procedure until no points are left. Then the recursion has come to an end.

It can be seen that the S-CH algorithm is quite simple. In the following experimental results will be presented.

## 3    Experimental Results

The proposed S-CH algorithm has been implemented in C# using .Net Framework 4.5 and tested on data sets using a PC with the following configuration:
- CPU: Intel® Core™ i7-2600 (4 × 3,40GHz)
- memory: 16 GB RAM
- operating system Microsoft Windows 7 64bits

### 3.1  Distribution of Points

The proposed S-CH algorithm has been tested using different $3D$ datasets. These datasets have different types of distributions of points. For experiments, we used well-known distributions such as randomly distributed uniform points in a unit sphere, uniform points in a unit cube, points lying on a unit sphere or points with a Gaussian distribution. Other distributions used were Halton points and Gauss ring points, which are described in the following text. Furthermore, we describe how to generate uniform spherical data.

**Spherical Points.** For generating uniform spherical points, spherical coordinates cannot be used, because these coordinates cause the points to be concentrated around poles. Therefore, we use the following approach to generate spherical points.

First, we generate a point $P$ lying in a cube, which represents an axis-aligned bounding box for a unit sphere, and determine the Euclidean norm of this point $\|P\|$. If $\|P\| > 1$, then we return to the start. Otherwise we normalize point $P$. Finally, we multiply this point by the required radius. The value of a radius can be either the same for all points (points on the sphere) or randomly generated for each point.

**Halton Points.** Construction of a Halton sequence is based on a deterministic method. This sequence generates well-spaced "draws" points from the interval $[0, 1]$. The sequence uses a prime number as its base and is constructed based on finer and finer prime-based divisions of sub-intervals of the unit interval. The Halton sequence [8] can be described by the following recurrence formula:

$$Halton(p)_k = \sum_{i=0}^{\lfloor \log_p k \rfloor} \frac{1}{p^{i+1}} \left( \left\lfloor \frac{k}{p^i} \right\rfloor \mod p \right), \tag{2}$$

where $p$ is the prime number and $k$ is the index of the calculated element.

For the $3D$ space, subsequent prime numbers are used as a base. In our test, we used {2,3,5} for the Halton sequence and we got a sequence of points in a unit cube:

$$\begin{aligned}Halton(2,3,5) = &\left\{ \left(\frac{1}{2},\frac{1}{3},\frac{1}{5}\right), \left(\frac{1}{4},\frac{2}{3},\frac{2}{5}\right), \left(\frac{3}{4},\frac{1}{9},\frac{3}{5}\right), \left(\frac{1}{8},\frac{4}{9},\frac{4}{5}\right), \left(\frac{5}{8},\frac{7}{9},\frac{1}{25}\right), \right.\\ &\left. \left(\frac{3}{8},\frac{2}{9},\frac{6}{25}\right), \left(\frac{7}{8},\frac{5}{9},\frac{11}{25}\right), \left(\frac{1}{16},\frac{8}{9},\frac{16}{25}\right), \left(\frac{9}{16},\frac{1}{27},\frac{21}{25}\right), \dots \right\}\end{aligned} \tag{3}$$

Visualization of the dataset with $10^4$ points of the Halton sequence from (3) can be seen in Fig. 6. We can see that the Halton sequence in $3D$ space covers this space more evenly than randomly distributed uniform points in the unit cube.

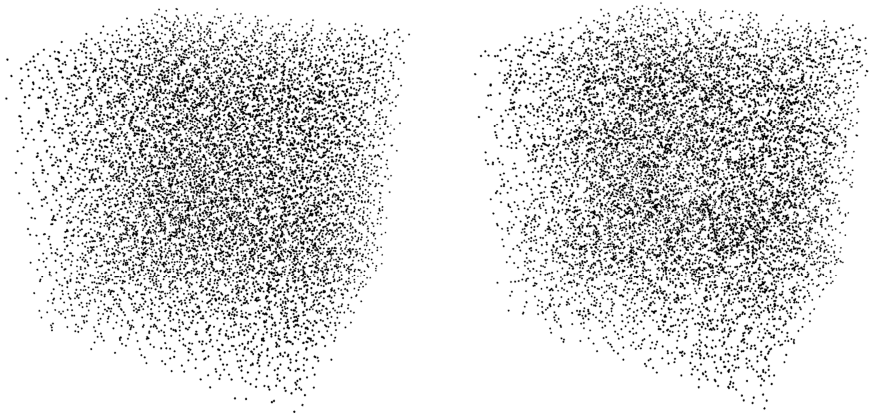

**Fig. 6.** $3D$ Halton points generated by $Halton(2,3,5)$ (left) and $3D$ random points in a cube with uniform distribution (right). Number of points is $10^4$ in both cases.

**Gauss Ring Points.** Construction of Gauss ring points in $3D$ space is based on the method for generating spherical points which is described above. For each point, the radius is determined using the following equation:

$$r = 0.5 + 0.5 \cdot sign \cdot rand_{Gauss}, \tag{4}$$

where $sign$ is a randomly generated number from set $\{-1,1\}$ and $rand_{Gauss}$ is a randomly generated number with Gauss distribution from interval $[0, \infty)$.

Visualization of the dataset with $10^4$ Gauss ring points can be seen in Fig. 7. We can see that this dataset consists of a large set of points, which are close to the sphere, and a small set of points, which are far from this sphere.

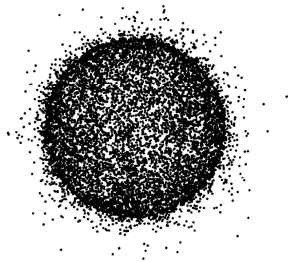

**Fig. 7.** $3D$ Gauss ring points. Number of points is $10^4$.

### 3.2 Examples of Convex Hull Generated

Some samples of convex hulls for datasets with a different distribution of points, which consist of $10^4$ points, are shown in Fig. 8.

It can be seen, the convex hull of points on a sphere or points with uniform distribution in a sphere has a spherical shape. Moreover, these convex hulls contain the majority of points. The convex hull of points with uniform distribution in a cube or Halton points is a box-shaped object. The random shape has a convex hull of Gauss points or Gauss ring points.

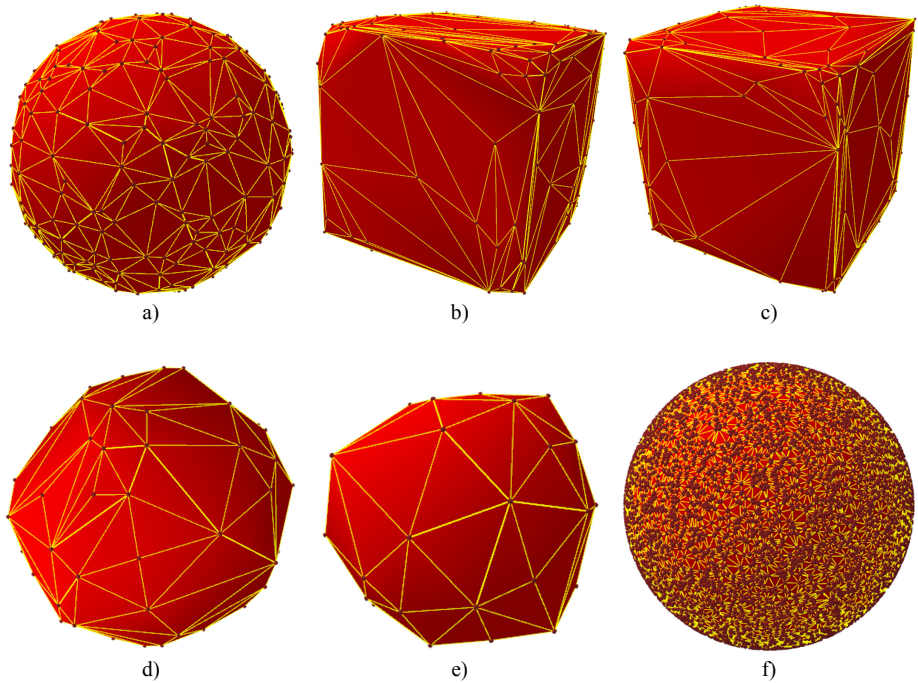

**Fig. 8.** Convex hulls of points with different distributions ($10^4$ points): a) uniform points in sphere, b) uniform points in cube, c) Halton points, d) Gauss points, e) Gauss ring points and f) points on sphere.

### 3.3 Optimal Number of Divisions

In the proposed approach, the main step is the division of the input set of points into non-overlapping sectors. Therefore, we need know an estimation of the optimal number of divisions, which should depend on the distribution of points. Consequently, we have to determine it for each type of input points separately.

We measured the time performance of the convex hull for different distributions of points, different numbers of points and different numbers of divisions. Measurement for $10^7$ points is presented in Graph 1. For all tested distributions of input points, except points on a sphere, we can see that the time performance decreases with the increasing number of divisions until the optimal number of divisions is achieved. After that time, the complexity increases with the increasing number of divisions. The situation is different for points on a sphere. Based on Graph 1f), it can be seen that the time complexity decreases with the increasing number of divisions. This is due to the fact that points are partially organized by the first step of the S-CH algorithm, and thus the construction of the final convex hull is accelerated. The speed up is gained due to better cache memory usage, more explained in [19].

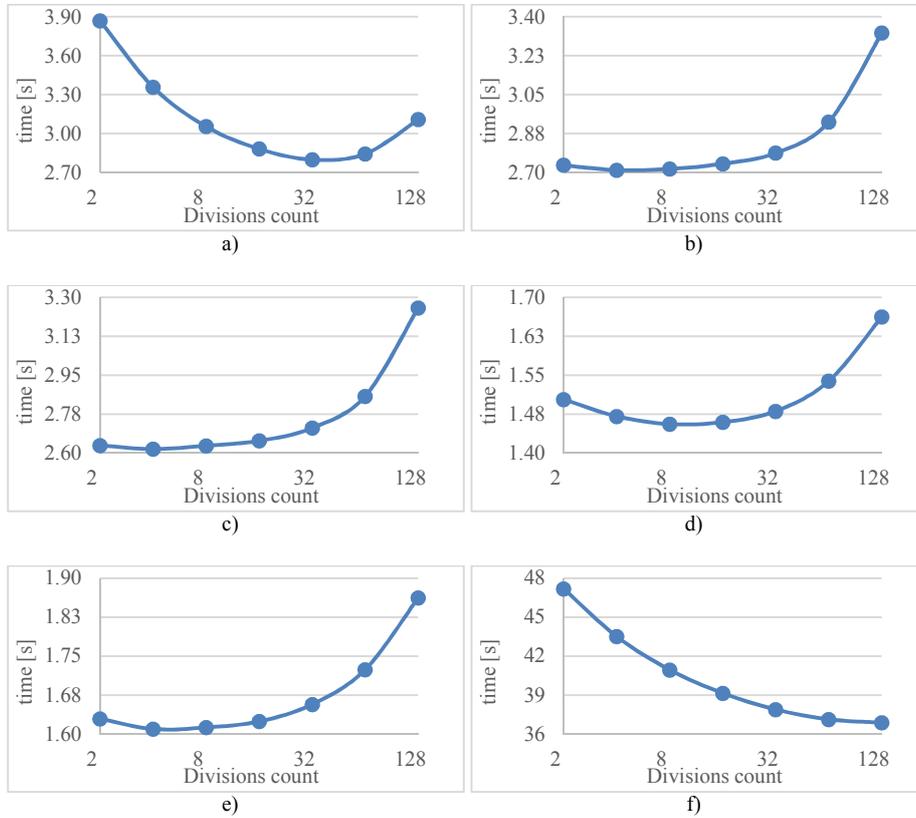

**Graph 1.** The time performance of the convex hull algorithm for different distributions of points and different division counts. The divisions count denotes the number of divisions in one axis, i.e. the total number of non-overlapping sectors is $6 \cdot (divisions\ count)^2$. The number of input points is $10^7$. Distributions of points are: a) uniform points in sphere, b) uniform points in cube, c) Halton points, d) Gauss points, e) Gauss ring points and f) points on sphere.

Evaluating experimental results for different numbers of input points, including results from Graph 1, i.e. $10^5$, $\sqrt{10} \cdot 10^5$, $10^6$, $\sqrt{10} \cdot 10^6$, $10^7$ and $\sqrt{10} \cdot 10^7$, we came to the following conclusion.

The expected optimal number of divisions is directly proportional to number of points lying on the convex hull. If the user knows properties of the input dataset, then the number of divisions can be determined more precisely. The optimal number of divisions, which is almost the same for all numbers of input points, is shown in Graph 2.

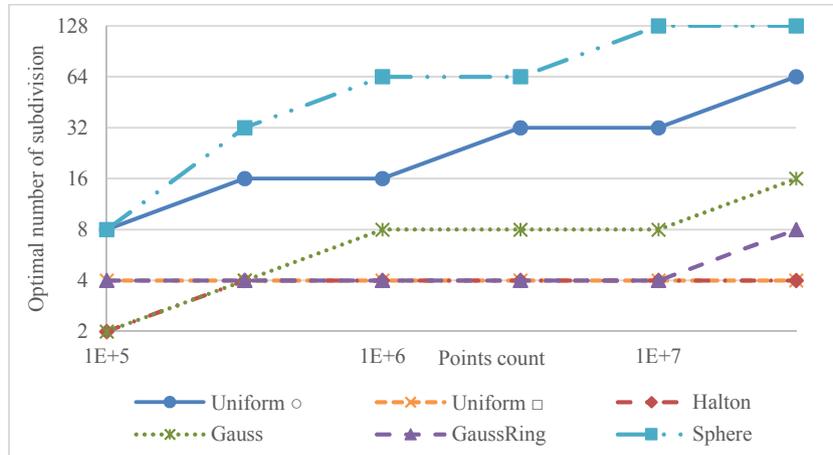

**Graph 2.** The optimal number of subdivisions for different numbers of input points and different distributions of these points.

### 3.4 Number of Points Processed at Each Step

In order to assess the effectiveness of the proposed algorithm, we need to know what proportion of input points to eliminate in each step of our algorithm, the size of the set of suspicious points and the number of points that lie on the convex hull. All these values are given relative to the size of the input dataset. Measurements were performed for different numbers of input points and different types of point distributions. The results of these experiments are in Table 2 - Table 6.

In Table 2 we can see the percentage of points eliminated by the initial polyhedron. It is obvious that the most points are eliminated for the Gauss distribution points. This is due to the fact that most of the Gauss points lie around the center. The number of points eliminated for Gauss ring points by the initial polyhedron is dependent on the total number of input points. From this, we can deduce, consistent with the Gauss ring distribution, that for smaller inputs, it may not be always possible to choose the ideal initial polyhedron. The results for the uniform distribution of points in a cube and Halton points are consistent with the theoretical estimate. (The theoretical estimate is obtained as the quotient of two volumes. The dividend is a volume of the ideal initial polyhedron and the divisor is a bounding volume for the input dataset.) The number of eliminated points is larger than the theoretical estimate for points with a uniform distribution inside a sphere.

**Table 2.** The percentage of points eliminated by the initial polyhedron.

| Number of points | Number of points eliminated [%] (100% means all the input data) | | | | | |
|---|---|---|---|---|---|---|
| | Uniform ○ | Uniform ☐ | Halton | Gauss | Gauss ⬤ | Sphere |
| 1E+5 | 49.37% | 12.77% | 13.71% | 89.79% | 47.65% | 0.00% |
| √10E+5 | 49.83% | 13.61% | 14.28% | 91.13% | 59.31% | 0.00% |
| 1E+6 | 50.16% | 15.22% | 14.27% | 91.88% | 67.67% | 0.00% |
| √10E+6 | 50.33% | 15.42% | 13.70% | 92.53% | 75.57% | 0.00% |
| 1E+7 | 50.41% | 15.14% | 14.91% | 92.85% | 82.24% | 0.00% |
| √10E+7 | 50.46% | 14.39% | 16.66% | 93.10% | 86.06% | 0.00% |

The percentage of points eliminated by the testing planes can be seen in Table 3. The most points are reduced for Halton points and for points with uniform distribution in a cube. For all tested distributions of input points, except points on a sphere, we can see that almost all input points are discarded after these two steps of the S-CH algorithm.

**Table 3.** The percentage of points eliminated by the testing planes.

| Number of points | Number of points eliminated [%] (100% means all the input data) | | | | | |
|---|---|---|---|---|---|---|
| | Uniform ○ | Uniform ☐ | Halton | Gauss | Gauss ⬤ | Sphere |
| 1E+5 | 40.42% | 72.22% | 72.64% | 6.62% | 43.11% | 0.00% |
| √10E+5 | 43.53% | 76.93% | 77.24% | 6.93% | 36.30% | 0.00% |
| 1E+6 | 45.07% | 78.87% | 80.29% | 7.10% | 30.11% | 0.00% |
| √10E+6 | 45.91% | 80.50% | 82.54% | 6.94% | 23.32% | 0.00% |
| 1E+7 | 46.51% | 82.01% | 82.50% | 6.86% | 17.18% | 0.00% |
| √10E+7 | 46.89% | 83.65% | 81.55% | 6.72% | 13.63% | 0.00% |

The percentage of points eliminated by reduction of suspicious points can be seen in Table 4. The minimal number of points is discarded by this step for all tested distributions of input points. But there exist distributions of points when this step is important, e.g. points forming a spiral. We can see non-zero values for points on a sphere. This is due to the elimination of initial points $R_i^{max}$, which were artificially added at the beginning (see Section 2.2) and lying on the faces of the initial polyhedron.

**Table 4.** The percentage of points eliminated by reduction of suspicious points.

| Number of points | Number of points eliminated [%] (100% means all the input data) | | | | | |
|---:|---:|---:|---:|---:|---:|---:|
| | Uniform ○ | Uniform □ | Halton | Gauss | Gauss ⬤ | Sphere |
| 1E+5 | 5.24% | 7.37% | 6.66% | 2.14% | 5.78% | 0.38% |
| $\sqrt{10}$E+5 | 2.93% | 4.24% | 3.84% | 1.22% | 2.82% | 0.12% |
| 1E+6 | 1.76% | 2.54% | 2.30% | 0.66% | 1.39% | 0.04% |
| $\sqrt{10}$E+6 | 1.13% | 1.63% | 1.49% | 0.35% | 0.72% | 0.01% |
| 1E+7 | 0.82% | 1.08% | 1.01% | 0.19% | 0.36% | 0.00% |
| $\sqrt{10}$E+7 | 0.60% | 0.74% | 0.69% | 0.11% | 0.19% | 0.00% |

The number of suspicious points for different numbers of input points and for different types of distributions is shown in Table 5. These points are used for the final calculation of the convex hull. It can be seen that for all tested distributions of points, except points on a sphere, the number of suspicious points is extremely low compared to the number of the original points.

**Table 5.** The percentage of suspicious points.

| Number of points | Number of candidates [%] (100% means all the input data) | | | | | |
|---:|---:|---:|---:|---:|---:|---:|
| | Uniform ○ | Uniform □ | Halton | Gauss | Gauss ⬤ | Sphere |
| 1E+5 | 5.36% | 8.03% | 7.38% | 1.83% | 3.84% | 100.00% |
| $\sqrt{10}$E+5 | 3.84% | 5.33% | 4.76% | 0.84% | 1.68% | 100.00% |
| 1E+6 | 3.05% | 3.41% | 3.18% | 0.39% | 0.86% | 100.00% |
| $\sqrt{10}$E+6 | 2.64% | 2.47% | 2.28% | 0.20% | 0.41% | 100.00% |
| 1E+7 | 2.27% | 1.77% | 1.59% | 0.11% | 0.22% | 100.00% |
| $\sqrt{10}$E+7 | 2.05% | 1.22% | 1.10% | 0.07% | 0.12% | 100.00% |

Table 6 presents the percentage of points lying on the final convex hull. Convex hulls of points with Gauss ring distribution, Gauss distribution, Halton distribution or uniform distribution in a cube are determined by the few remaining points. More points lie on the convex hull of uniform points in a sphere. The convex hull of points on a sphere should be determined by all these points, but the experimental results do not correspond to this assumption. This is due to the floating point precision of calculation.

**Table 6.** The percentage of points lying on the convex hull.

| Number of points | Number of points on the convex hull [%] (100% means all the input data) | | | | | |
|---|---|---|---|---|---|---|
| | Uniform ○ | Uniform □ | Halton | Gauss | Gauss ● | Sphere |
| 1E+5 | 1.100% | 0.191% | 0.183% | 0.173% | 0.086% | 95.016% |
| $\sqrt{10}$E+5 | 0.620% | 0.073% | 0.068% | 0.080% | 0.034% | 76.890% |
| 1E+6 | 0.350% | 0.027% | 0.025% | 0.038% | 0.013% | 42.410% |
| $\sqrt{10}$E+6 | 0.190% | 0.009% | 0.009% | 0.017% | 0.005% | 16.370% |
| 1E+7 | 0.110% | 0.003% | 0.003% | 0.008% | 0.002% | 5.541% |
| $\sqrt{10}$E+7 | 0.060% | 0.001% | 0.001% | 0.004% | 0.001% | 1.805% |

Moreover, we can see the percentage of suspicious points, which lie on the convex hull, in Graph 3.

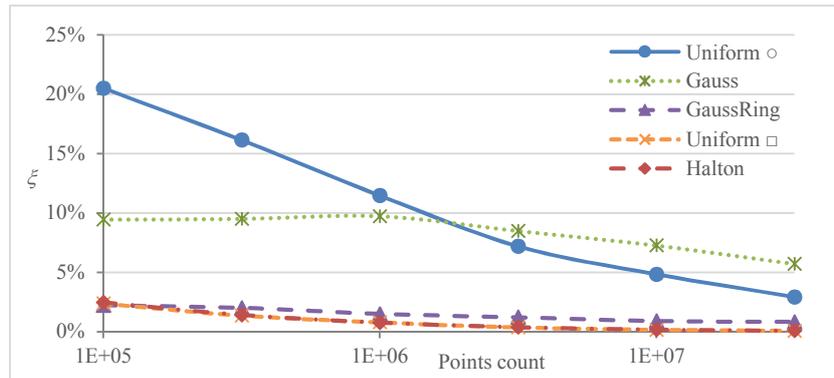

**Graph 3.** The percentage of suspicious points lying on the final convex hull for different numbers of input points and different distributions of these points.

### 3.5 Time Performance

In this section, we focus on running times for the calculation of a convex hull using our proposed S-CH algorithm. Running times were measured for different numbers of input points with different distributions of points. Measurements were performed many times and average running times, calculated from the measured results, are in Table 7 - Table 9; we can see these running times in Graph 4.

It can be seen that the best time performance is for datasets with the Gaussian distribution. These datasets are followed by Gauss ring points. This is expected behavior because most of the points using one of these distributions lie inside the initial polyhedron. Therefore, there are only a few points on the convex hull. The time performance for Halton points and for uniform points in a cube is similar. The running times for points with uniform distribution inside a sphere are a bit slower than the running times for uniform points in a cube. The worst time performance was obtained for points, which lie on a sphere. This is again expected behavior because there are no points for elimination during the first phase and therefore the convex hull calculation has to be done from the whole dataset.

**Table 7.** The time performance of the convex hull for different numbers of input points and different distributions of points. The number of divisions is equal to 4.

|  | Time [ms] | | | | | |
| --- | --- | --- | --- | --- | --- | --- |
| Number of points | Uniform ○ | Uniform □ | Halton | Gauss | Gauss ● | Sphere |
| 1E+5 | 36.6 | 30.7 | 30.3 | 15.5 | 23.9 | 868.0 |
| $\sqrt{10}$E+5 | 111.9 | 92.9 | 89.9 | 47.9 | 67.8 | 2 470.3 |
| 1E+6 | 339.0 | 284.2 | 274.4 | 149.5 | 189.2 | 5 857.8 |
| $\sqrt{10}$E+6 | 1 057.8 | 875.8 | 849.2 | 465.5 | 554.3 | 14 426.4 |
| 1E+7 | 3 357.6 | 2 710.8 | 2 619.1 | 1 471.2 | 1 609.4 | 43 515.7 |
| $\sqrt{10}$E+7 | 10 792.3 | 8 497.7 | 8 192.9 | 4 611.7 | 4 898.7 | 160 238.9 |

**Table 8.** The time performance of the convex hull for different numbers of input points and different distributions of points. The number of divisions is equal to 8.

|  | Time [ms] | | | | | |
| --- | --- | --- | --- | --- | --- | --- |
| Number of points | Uniform ○ | Uniform □ | Halton | Gauss | Gauss ● | Sphere |
| 1E+5 | 35.1 | 31.6 | 31.2 | 16.1 | 24.9 | 865.6 |
| $\sqrt{10}$E+5 | 104.4 | 94.1 | 91.1 | 48.6 | 68.8 | 2 451.1 |
| 1E+6 | 312.1 | 285.1 | 277.5 | 149.4 | 189.3 | 5 760.5 |
| $\sqrt{10}$E+6 | 970.8 | 879.9 | 849.7 | 463.1 | 554.9 | 13 989.2 |
| 1E+7 | 3 054.4 | 2 716.4 | 2 633.6 | 1 456.2 | 1 612.5 | 40 942.3 |
| $\sqrt{10}$E+7 | 9 715.4 | 8 541.4 | 8 215.8 | 4 567.2 | 4 877.4 | 146 343.2 |

You can see the best average running time for the optimal number of divisions for each distributions of points and different numbers of input points in Table 9.

**Table 9.** The time performance of the convex hull for different numbers of input points and different distributions of points. The results are presented for the optimal number of divisions, see Graph 2.

|  | Time [ms] | | | | | |
| --- | --- | --- | --- | --- | --- | --- |
| Number of points | Uniform ○ | Uniform □ | Halton | Gauss | Gauss ● | Sphere |
| 1E+5 | 35.1 | 30.7 | 30.2 | 15.5 | 23.9 | 865.6 |
| $\sqrt{10}$E+5 | 103.4 | 92.9 | 89.9 | 47.9 | 67.8 | 2 436.4 |
| 1E+6 | 300.5 | 284.2 | 274.4 | 149.4 | 189.2 | 5 666.3 |
| $\sqrt{10}$E+6 | 910.8 | 875.8 | 849.2 | 463.1 | 554.3 | 13 358.2 |
| 1E+7 | 2 799.4 | 2 710.8 | 2 619.1 | 1 456.2 | 1 609.4 | 36 873.2 |
| $\sqrt{10}$E+7 | 8 718.4 | 8 497.7 | 8 192.9 | 4 564.4 | 4 877.4 | 121 323.3 |

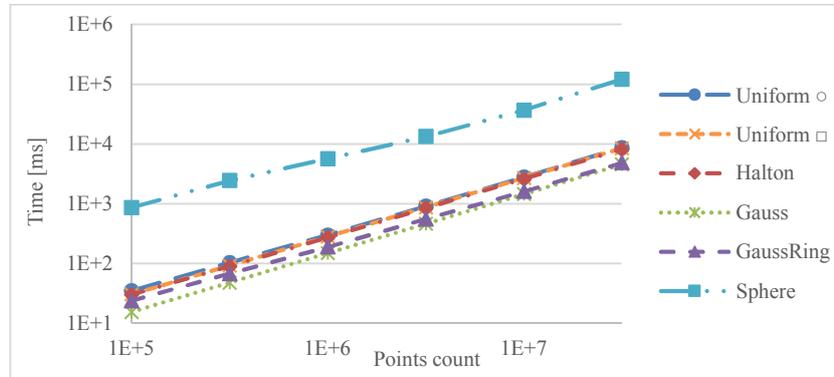

**Graph 4.** The time performance of the convex hull for different numbers of input points and different distributions of these points.

Moreover, we were performed the measurements for different real-world examples, see Fig. 9. Average running times for S-CH algorithm and QuickHull algorithm are in Table 10. We can see that our proposed S-CH algorithm give for dataset of MRI of brain better time performance than QuickHull algorithm. Contrary, time performance of QuickHull algorithm is better than S-CH algorithm for dataset of laser scanned bunny.

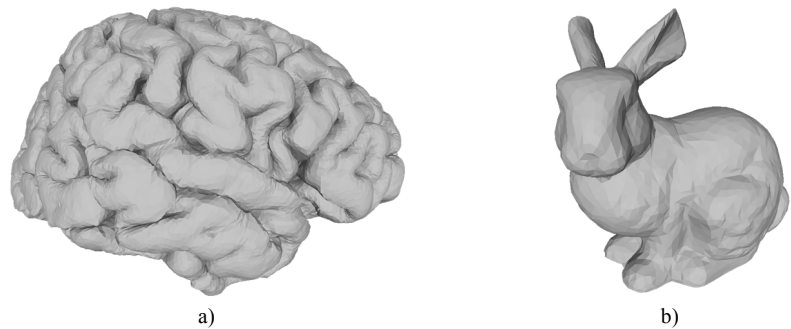

a)        b)

**Fig. 9.** Input sets for convex hull computation: a) MRI of brain (9 247 234 points), b) laser scanned bunny (35 947 points).

**Table 10.** The time performance of the convex hull for different real-world examples. The optimal number of divisions is equal to 16.

| Model | Number of points | Time [ms] | |
|---|---|---|---|
| | | S-CH algorithm | QuickHull |
| bunny | 35 947 | 62.0 | 47.0 |
| brain | 9 247 234 | 2925.0 | 4 118.5 |

### 3.6 Comparison with Other Algorithms

We compared the proposed S-CH algorithm with the incremental insertion algorithm and QuickHull algorithm, whose expected time complexity is $O(N \log N)$, and with the Chan's algorithm, which expected time complexity is $O(N \log h)$, where $N$ is the number of input points and $h$ is the number of points on the output convex hull. It should be noted that we use the library MIConvexHull, which is implemented in C# using .Net Framework 4.5, for measurements of the QuickHull algorithm. The results for the incremental insertion algorithm are based on the use of the ratio of the Randomized Incremental algorithm to QuickHull. This ratio was obtained from measurements for a C implementation of both algorithms.

Running times were measured for different numbers of input points with uniform distribution inside a sphere. The resultant speed-up of the S-CH algorithm with respect to the QuickHull algorithm, Chan's algorithm and Randomized Incremental algorithm can be seen in Graph 5.

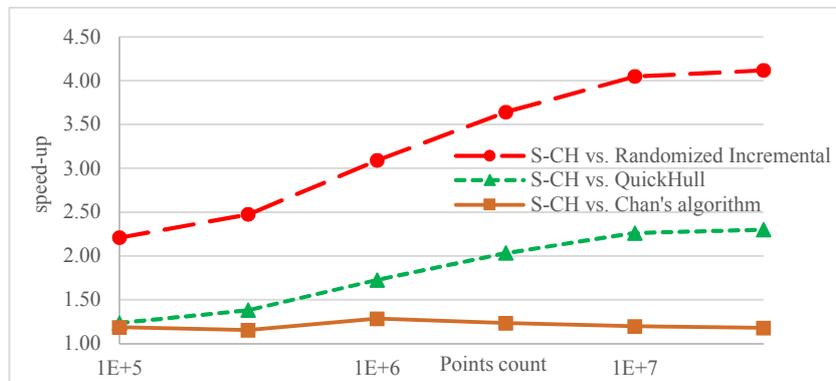

**Graph 5.** The speed-up of the S-CH algorithm for points in a sphere with uniform distribution with respect to QuickHull, Chan's algorithm and Randomized Incremental algorithm for the same datasets.

It can be seen that the proposed S-CH algorithm clearly outperforms "standard" convex hull algorithms. The graph shows that speed-up grows slowly from $10^7$ points. This is due to swapping.

## 4   Conclusion

A new fast convex hull algorithm in $E^3$ has been presented. The S-CH algorithm uses a space division technique. It is robust, as we do not use any angle calculations, and can process a large number of points as well as different distributions of points. Advantages of the S-CH algorithm are simple implementation, robustness and the use of almost any known algorithm for the final calculation of the convex hull as very efficient filtering, very small number of points are left for the final processing, i.e. for the final convex hull construction. Therefore the final efficiency is not sensitive to the convex hull algorithm properties. Thus, any brute force algorithm, which is easy to implement and robust, can also be used without significant influence to the algorithm efficiency. We do not assume any special order of input points. Otherwise, there is a possibility for a modification to increase the effectiveness of the algorithm.

For future work, the S-CH algorithm can be easily parallelized, as most of the steps are independent, and for large datasets influence of caching and data transfer should be explored more deeply.

**Acknowledgments.** The authors would like to thank their colleagues at the University of West Bohemia, Plzen, for their comments and suggestions, and anonymous reviewers for their valuable comments and hints provided. The research was supported by MSMT CR projects LH12181 and SGS 2013-029.